\documentclass[prb,  twocolumn, showpacs, superscriptaddress,longbibliography]{revtex4-1}
\pdfoutput=1
\usepackage{amsmath, amsfonts, amssymb, mathrsfs}
\usepackage{graphicx, color}

\usepackage[caption=false]{subfig}
\usepackage{bm}
\usepackage{braket}
\usepackage{epstopdf}
\usepackage{dsfont}
\usepackage{ulem} 
\usepackage[dvipsnames]{xcolor}
\usepackage{program}
\usepackage[hyperindex]{hyperref}
\hypersetup{
   colorlinks,
   citecolor=blue,
   filecolor=blue,
   linkcolor=blue,
   urlcolor=blue
}
\usepackage{chngcntr}

\graphicspath{Graphes}

\renewcommand{\underline}[1]{\uline{#1}}
\newcommand{\be}{\begin{equation}}
\newcommand{\ee}{\end{equation}}

\newcommand{\Aa}{\mathcal{A}}
\newcommand{\Bb}{\mathcal{B}}

\newcommand{\Ss}{\mathcal{S}}

\definecolor{orange}{rgb}{1,0.5,0}

\definecolor{grey}{rgb}{.6,.6,.6}

\begin{document}

\title{Multiscale entanglement clusters at the many-body localization phase transition}

%
\author{Lo\"{i}c Herviou}
\affiliation{Department of Physics, KTH Royal Institute of Technology, Stockholm, 106 91 Sweden}
\author{Soumya Bera}
\affiliation{Department of Physics, Indian Institute of Technology Bombay, Mumbai 400076, India}
\author{Jens H.~Bardarson}
\affiliation{Department of Physics, KTH Royal Institute of Technology, Stockholm, 106 91 Sweden}

\begin{abstract}
We study numerically the formation of entanglement clusters across the many-body localization phase transition. 
We observe a crossover from strong many-body entanglement in the ergodic phase to weak local correlations in the localized phase, with contiguous clusters throughout the phase diagram.
Critical states close to the transition have a structure compatible with fractal or multiscale-entangled states, characterized by entanglement at multiple levels: small strongly entangled clusters are weakly entangled together to form larger clusters.  
The critical point therefore features subthermal entanglement and a power-law distributed cluster size, while the localized phase presents an exponentially decreasing cluster distribution.
These results are consistent with some of the recently proposed phenomenological renormalization-group schemes characterizing the many-body localized critical point, and may serve to constrain other such schemes. 
\end{abstract}
\date{\today}
\maketitle

\section{Introduction}
Entanglement has been central in developing our current understanding of many-body localization\citep{Gornyi2005, Basko2006, Abanin2017, Alet2018}.
The many-body localization transition is fruitfully viewed as an eigenstate phase transition with eigenstate transitioning from area law entanglement in the localized phase at strong disorder\citep{Bauer2013, Kjall2014} to a volume law entanglement in the ergodic phase at weak disorder\citep{Luitz:2016hr,Yu:2016gb}.
The area law is a direct consequence of an emergent integrability in the localized phase\citep{Serbyn2013, Huse2014,Imbrie2016, Imbrie2016-2} and eigenstates being product states of local integrals of motion\citep{Serbyn2013, Huse2014, Chandran2015, Ros2015, Rademaker2016, Bera2015, Bera2016, OBrien2016, Lezama2017}.
This also results in the experimentally observed absence of thermalization\citep{Schreiber2015, Choi2016, Luschen2017}, and slow growth of entanglement in quenches follows from the exponentially decaying interactions between the localized integrals of motion\citep{Znidaric2008,Bardarson2012,Serbyn2013a}.
The ergodic phase, in turn, exhibits slow dynamics\citep{Agarwal2017,Luitz:2017cp}, at least at small system sizes and short times, even in periodically driven systems~\citep{Roy:2018bb,Lezama:2018ti}, with a subballistic growth of entanglement.
The subballistic entanglement growth is observed to slow down when approaching the transition and turns, at the critical point, into a logarithmic growth\citep{Znidaric2008,Bardarson2012}.

When it comes to the exact nature of the MBL phase transition the picture is less clear, and the transition has proven resilient to both numerical and analytical approaches. 
As a result, several phenomenological renormalization group (RG) approaches have been proposed\citep{Potter2015, Vosk2015, Zhang2016, Monthus2017, Dumitrescu2017, Thiery2017, Thiery-2017-long, Goremykina2018, Igloi2018, Dumitrescu2018}, based on various physical assumptions on thermalization and random matrix theory. 
These approaches paint a different, albeit similar, picture of the critical point, where thermal and localized \textit{clusters} of spins alternate or mingle. 
Here a cluster is to be understood as a subset of spins that is significantly less entangled with the rest of the system than within itself. 
The appearance of these clusters can be understood as a precursor of the emergent integrability and the local integrals of motions\citep{Serbyn2013, Huse2014, Chandran2015, Ros2015, Rademaker2016, Imbrie2016, Imbrie2016-2, OBrien2016}  characteristic of the MBL phase. Indeed, a completely disentangled cluster suggests the existence  of conserved quantities, such as its total charge.

\begin{figure}
\includegraphics[width=1\columnwidth]{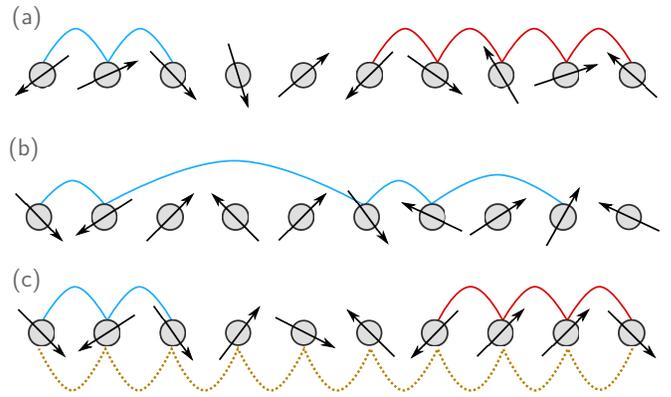}
\caption{Schematics for different types of entanglement structures obtained in different RG proposals. 
Black arrows represent spins, while the solid links mark entanglement between them. 
A set of spins connected by these links forms a single ergodic cluster, with different colors associated to different clusters. 
MBL clusters correspond to continuous sets of unconnected spins. 
The dotted lines mark a weaker entanglement. 
Though some residual entanglement survives in a MBL cluster, for simplicity, it is not represented here. 
(a) Ergodic and MBL connected clusters simply alternate.  
(b) One sparse ergodic cluster spans the entire chain. 
(c) Multi\-scale entanglement with small strongly entangled clusters (full lines) weakly entangled at a higher scale (dotted line) to form a larger cluster.
}
\label{fig:clusters-RG}
\end{figure}

A distinguishing feature of the different RG proposals is the predicted structure of entanglement in eigenstates at the critical point.
Three different proposals are depicted in Fig.~\ref{fig:clusters-RG}:
the first has alternating ergodic and localized clusters, each consisting of a consecutive sequence of spins\citep{Vosk2015};
the second suggests a dilute ergodic cluster that spans the entire chain, while most of the spins are localized\citep{Potter2015, Khemani2017};
the third multiscale (or fractal) state has a layered structure, with small ergodic and MBL clusters joining up at a larger scale to form bigger clusters\citep{Zhang2016, Thiery2017, Thiery-2017-long, Goremykina2018, Dumitrescu2018}.
To determine which of these different scenarios best corresponds to the actual physics requires identifying and analyzing the entanglement structure of eigenstates obtained from microscopic models---an algorithm for such identification is one of the main results of this work.

Thermal and many-body localized clusters can be identified by considering their entanglement properties.
Indeed, in the ergodic phase, for a subsystem $\Aa$ of $n_\Aa$ spins, the (bipartite) von Neumann entanglement entropy
\begin{equation}
\Ss(\Aa)=- \text{Tr } \rho_\Aa \ln \rho_\Aa, 
\end{equation}
where $\rho_\Aa$ is the reduced density matrix describing $\Aa$, takes a thermal value that depends on the energy density of the state and is linear in $l_\Aa$\citep{Page1993, DAlessio2016}. 
For middle of spectrum states, this thermal value is $n_\Aa \ln 2$ in the thermodynamic limit. 
The internal entanglement of the thermal clusters also follows this volume law, albeit with a possibly lower coefficient, due to finite-size effects.
On the other hand, many-body localized systems (and similarly, localized clusters) admit a lower entanglement entropy that follows an area law~\cite{Bauer2013}.

In this work, we study numerically the formation, properties and structure of clusters in a typical one-dimensional spin system, the XXZ spin chain in a random transverse field. 
Our method relies on an entanglement-based decomposition of the wave function and can be easily extended to more general models, including fermionic systems.
Using a recursive search algorithm, we first identify the entanglement structure in the ergodic and localized phases. 
Entanglement in the ergodic phase is spread out over the entire system, without any local structure. 
In the localized phase, it is weaker and local: rare strongly entangled pairs survive even at strong disorder.
The critical point can be identified by considering the characteristic sizes (number of spins and range) of the clusters. 
We show that the critical states admit multiscale entanglement that leads to a hierarchy of contiguous clusters. 
This leads to subthermal entanglement and a power-law distributed cluster size at the critical point, in qualitative agreement with recently proposed RG schemes\citep{Dumitrescu2017, Thiery2017, Thiery-2017-long, Goremykina2018, Dumitrescu2018}.

\section{Entanglement cluster identification and characterization}
Ideally, to identify entanglement clusters, we should consider all possible partitions of the system and identify the one minimizing some appropriate multipartite entanglement measure. 
However, the factorial number of partitions and the general (at least) exponential cost of such entanglement measures make this approach numerically infeasible, except for the very smallest of system sizes.
Instead, we introduce an approximate recursive algorithm, restricting ourselves only to bipartite measurements.
To quantify the entanglement between two subparts $\Aa$ and $\Bb$ of an ensemble $\Aa\cup\mathcal{B}$, we use the normalized mutual information:
\begin{equation}
0 \leq i(\Aa, \mathcal{B})=\frac{\Ss(\Aa)+\Ss(\mathcal{B})-\Ss(\Aa\cup\mathcal{B})}{\min( n_\Aa ,n_{\mathcal{B}})} \leq 2 \ln 2,
\end{equation}
with $n_\Aa$ and $n_{\mathcal{B}}$ the number of spins in clusters $\Aa$ and $\mathcal{B}$ respectively.
When $\Aa\cup\mathcal{B}$ is the total system, $i(\Aa,\mathcal{B})$ reduces to twice the normalized entanglement entropy. 
Mutual information is a reliable tool to study the MBL phase transition as it distinguishes between wave functions with many-body entanglement,  strongly-entangled pairs of spins, or localized spins\citep{DeTomasi2017, Magan2017, Banuls2017, Iyoda2018, Kjall2018}.

\begin{figure}
\includegraphics[width=0.8\linewidth]{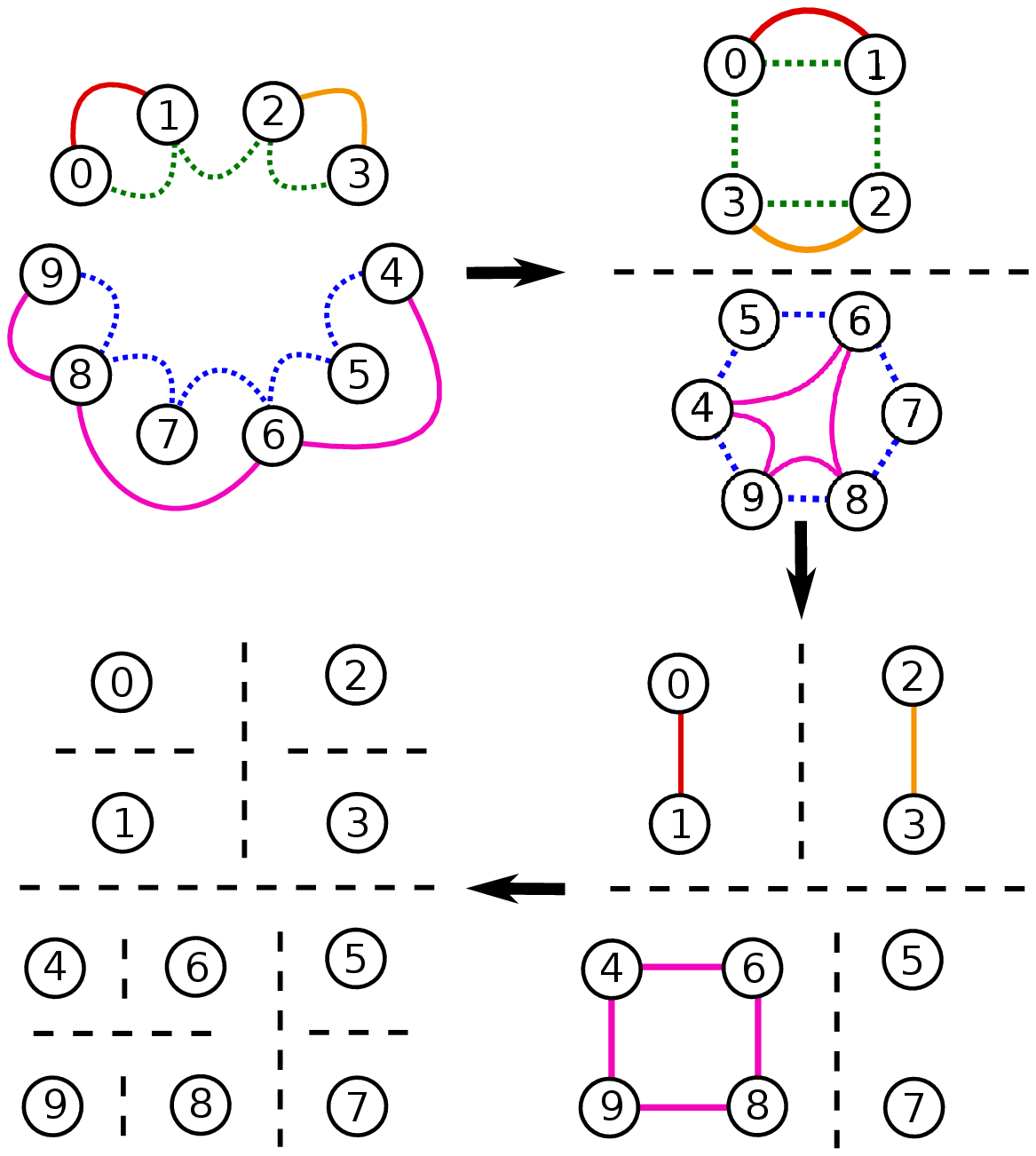}
\caption{Schematic description of the recursive algorithm used to identify entanglement clusters. 
The indexed circles represent physical sites, while links illustrate entanglement between sites. 
Dotted lines represent weaker correlations and the dashed lines the splitting into different clusters. 
The arrows mark one step of the algorithm, wherein the continuous bipartition that minimizes the mutual information is identified. 
We then apply the same process recursively to each so-obtained cluster.}
\label{fig:Algo}
\end{figure}

As a first step towards identifying clusters, we represent the wave function as a binary tree with each node representing a set of spins.
The root of the tree consist of all spins in the system.
The first decomposition into two sets is identified by the bipartition of the system that minimizes the normalized mutual information.
We then recursively search for the optimal decomposition of each set into two parts, until all sets are single-site, as illustrated in Fig.~\ref{fig:Algo}. 
We define the internal mutual information of a set as the normalized mutual information of its two descendants, i.e., the minimum amount of entanglement between all its bipartitions; for a single-site set, we take the internal mutual information to be its maximal possible value $2 \ln 2$. 

As the number of bipartitions of a system still scales exponentially, we further restrict the decompositions we evaluate:
at each step of the algorithm, we only consider bipartitions that would be continuous if the set we divide was an isolated system with periodic boundary conditions.
To give a concrete example, if the current set consisted of sites $\{1, 2, 7, 8, 9\}$, $\{ 1, 8, 9\}$ would be a potential subset, while $\{1, 7, 9\}$ would not.
This gives us the possibility to describe sparse sets at a low numerical cost, while retaining sufficient descriptive power to distinguish all the relevant structures observed in this paper.
The normalization of the mutual information favors partitioning large localized sets before removing single spins from thermal sets or breaking strongly-entangled pairs.

The second step is to extract the cluster structure of the wave function from the binary tree.
To do so, we introduce a mutual information cut-off $i_{\text{cut}}$. 
Starting from the root, that is to say the whole system considered as a single cluster, we apply recursively the following rule: sets whose internal normalized mutual information is larger than this cut-off are kept as clusters, while the others are divided into their two children. 
The final stable ensemble of clusters is the cluster decomposition at the cut-off $i_{\text{cut}}$.
Varying $i_{\text{cut}}$ reveals the details of the entanglement structure of a state:
at $i_{\text{cut}}=0$, the full system is considered to be a single cluster; 
with increasing $i_{\text{cut}}$, independent clusters progressively appear and are then broken down eventually into single-site sets, starting with the weakly entangled localized clusters.
We further introduce a \textit{minimal clustering}, defined by fixing the entanglement cut-off in each state to be equal to the internal mutual information obtained at the first partition of our algorithm, denoted $i_{\min}$.
Note that the minimal clustering is not necessarily a simple bipartition of the total system: if one of the sets obtained after the first bipartition has smaller internal entanglement than the initial complete system, it will be further broken down.
The minimal clustering gives additional information on the way entanglement is spread through the system.
In particular, it differentiates between many-body entanglement, with entanglement shared equally between all spins in a cluster, and resonant mechanisms where a spin couples with a single or few other spins.

For a given cluster $\Aa=\{x_1,..., x_{n}\}$ of $n$ spins, with  $x_1<x_2<...<x_{n}$, we define the range of the cluster 
\begin{equation} 
l = 1+\min_{j=1,..., n}(x_j-x_{j+1}\mod L)
\end{equation}%
with $x_{n+1}=x_1$.
While $n$ is simply the number of sites in the cluster, $l$ corresponds to the range of the system over which the cluster spreads. 
The spread  controls the decay of correlation functions, as distant spins have nonzero long-range correlations if they are in the same cluster.
The difference $l-n$ is a simple marker of the sparsity of a given cluster.
Denoting with $[\ . \  ]$ the average over clusters in a given state and $\langle\ . \ \rangle$ the average over states and disorder realizations, we define the average number of spins in a cluster $n_\mathrm{av} = \langle [ n ] \rangle$, and the average maximum size of a cluster in chain $n_\mathrm{max} = \langle\max\ n\rangle$.
Analogous definitions hold for the range averages $l_\mathrm{av}$ and $l_\mathrm{max}$.

We  verified that our algorithm distinguishes between all entanglement structures described in Fig.~\ref{fig:clusters-RG}, by applying it to randomly generated wave functions; the results are shown in Appendix \ref{app-Validity}.
Only intertwined thermal clusters, such as $\{1, 3, 5\}$ and $\{2, 4\}$ in a $L=5$ chain, cannot be clearly identified, due to the connectivity approximation (the structure  $\{1, 3, 5\}$,  $\{2\}$, and $\{4\}$, on the other hand, is readily identified).
One would generally expect such clusters to merge into a single cluster in a physical system, so this approximation does not seem limiting.
In small systems, where all bipartitions can be computed, we indeed verify that we obtain identical results when relaxing the assumption of continuous clusters; see Fig.~\ref{fig:ComparisonComplete} in Appendix~\ref{app-Validity}.

\section{Cluster structure of the ergodic and MBL phases}
We apply our algorithm to the eigenstates of the XXZ spin chain in a transverse field, a common model of many-body localization\citep{Znidaric2008, Pal2010, Bauer2013, Serbyn2013, DeLuca2013, Nanduri2014, Luitz2015}, with Hamiltonian 
\begin{equation}
H=- \sum\limits_{j} (\sigma^+_j \sigma^-_{j+1}  + \sigma^-_j \sigma^+_{j+1}) + \frac{\Delta}{2} \sigma^z_j \sigma^z_{j+1} + h_j\sigma^z,\label{eq:Hamiltonian}
\end{equation}
where  $\sigma$ are the Pauli matrices, $\Delta$ controls the interaction strength, and $h_j$ is a random magnetization. 
We take periodic boundary conditions and denote with $L$ the number of sites ranging from $10$ to $18$. 
Unless otherwise specified, we take $\Delta=1$ and $h_j^z$ uniformly distributed in $[-W, W]$.
This Heisenberg limit admits a phase transition from an ergodic (metallic) phase to a many-body localized phase.
The critical value of disorder in exact diagonalization studies of up to $L=26$ spins is obtained as $W_c \approx 3.6$\citep{Berkelbach2010, Pal2010, Luitz2015, Pietracaprina2017, Pietracaprina2018}, while some matrix product state or Schmidt gap based approaches argue for somewhat larger critical disorder strength~\citep{Devakul2015, Gray2018, Doggen2018}.
We compute through exact diagonalization the 50 eigenstates with energies closest to the middle of the spectrum for 1000-5000 disorder realizations.

\begin{figure}
\begin{center}
\includegraphics[width=1\columnwidth]{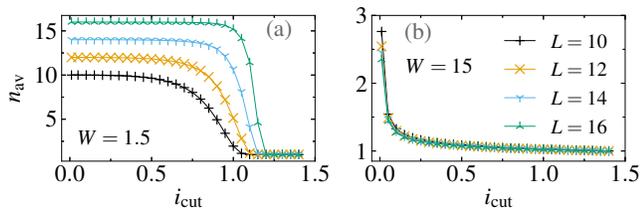}
\end{center}	
\caption{Average number of sites in a cluster as a function of mutual information cut-off for different system sizes for $W=1.5$ (a) and $W=15$ (b). 
For $W=1.5$, the system is ergodic and below the thermal value of $i_\mathrm{cut}$, which is renormalized by finite-size, the complete system forms a single cluster; above this value, clusters become single-site. 
The abrupt transition marks true many-body entanglement and is a sign of ergodicity. 
For $W=15$, the system is many-body localized. 
Even at low values of the entanglement cut-off, the system is fragmented into smaller clusters. 
With increasing mutual information cut-off, these clusters become smaller, with some resonating pairs surviving even at high cut-off.}
\label{fig:avLen-1}
\end{figure}

Deep in the ergodic phase, the eigenstates have volume-law entanglement, shared equally between all sites. 
Consequently, after the first partition, the internal entanglement within each cluster will be smaller than $i_{\text{min}}$, and the minimal clustering corresponds to single-site clusters. 
Similarly, as long as  $i_{\text{cut}}$ is below the thermal value, our algorithm considers the chain as a single cluster. 
This behaviour is observed in Fig. \ref{fig:avLen-1}\hyperref[fig:avLen-1]{(a)}. 
This transition from global to single-site clusters marks truly many-body entanglement and is characteristic of the ergodic phase in general.

\begin{figure}[tb!]
\begin{center}
\includegraphics[width=1\columnwidth]{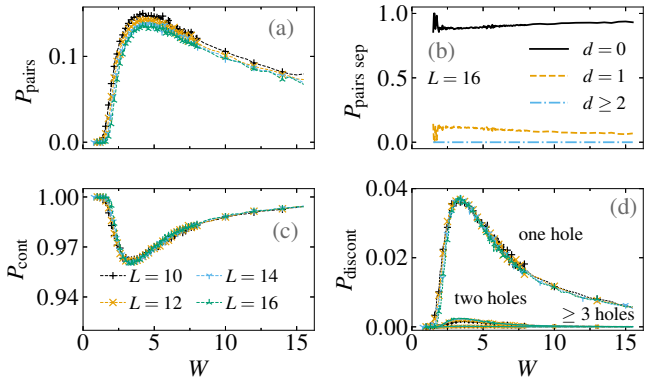}
\end{center}
\caption{(a) Fraction of clusters that consist of a pair of spins, and (b) fraction of pairs that are separated by $d$ sites for $L=16$. 
Resonant pairs are still present at high disorder strength (while the fraction of pairs decreases with system size, the number of clusters increases faster), but are short range with at most next-nearest neighbours.
(c) Fraction of continuous clusters as a function of disorder strength. 
(d) Fraction of clusters with holes (here taken to be $l-n$); there are up to $4\%$ with a single hole, less than $1\%$ with two holes, and practically no clusters with a larger number of discontinuities. 
All data are obtained for $i_\text{cut}=0.35$. 
}
\label{fig:Continuity}
\end{figure}

In the MBL phase, the spins are very weakly entangled and $\langle{i_{\min}}\rangle$  approaches zero as disorder increases. 
As we normalize the mutual information by the number of sites, minimal clusters have a length close to $L/2$. 
Some rare strongly entangled pairs are still present, with seemingly finite probability in the thermodynamic limit. 
Figure \ref{fig:Continuity}\hyperref[fig:Continuity]{(a-b)} shows the probability $P_\mathrm{pairs}$ for a cluster to consist of a pair of spins, and the probability $P_\text{pairs sep}(d)$ for the two spins in a pair to be separated by a distance $d$.
We only observe short range pairs: for $W\geq 6$, around $90\%$ of pairs correspond to neighbouring sites, while around $10\%$ are separated by a single site. 
These ratios depend only weakly on $i_\mathrm{cut}$, and we do not observe further separated pairs in any of our samples.
The continuity of the measured clusters is presented in Fig.~\ref{fig:Continuity}\hyperref[fig:Continuity]{(c-d)}.
In both phases, the difference between the range of a cluster and its number of sites is negligible.
The probability $P_\mathrm{discont}$ for a cluster to be discontinuous peaks close to the phase transition, but remains small (less than $5\%$ for $i_\mathrm{cut}=0.35$). 
We essentially never observe a discrepancy between $l$ and $n$ larger than $2$, at any cut-off. 

\section{Cluster structure at the critical point}
\begin{figure}
\begin{center}
\includegraphics[width=1\columnwidth]{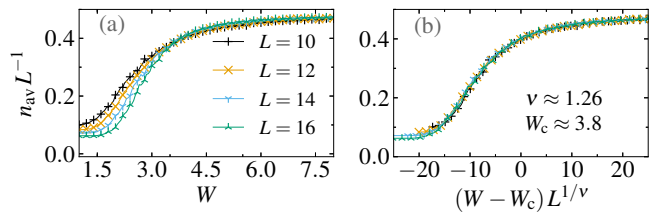}
\end{center}	
\caption{$(a)$ Mean length of minimal clusters as a function of disorder strength and $(b)$ scaling collapse to the universal scaling form $n_\mathrm{av}=L^{-1} f((W-W_c)L^{1/\nu})$. 
Similar results are obtained for the mean range. 
The collapse suggests that these quantities act as good order parameters for the  MBL phase transition.
}
\label{fig:avLen-2}
\end{figure}

We plot in Fig.~\ref{fig:avLen-2}\hyperref[fig:avLen-2]{(a)}  the average length $n_\mathrm{av}$ of the minimal clusters, as a function of disorder strength. 
Both $n_\mathrm{av}$ and the average range $l_\mathrm{av}$ act as good scaling order parameters for pinpointing the ergodic to MBL phase transition, as they collapse remarkably well around the known critical point if rescaled by the total system length. 
The value of the critical disorder $W_c\approx 3.8$ is in good agreement with earlier studies, and the critical exponent is $\nu = 1.26\pm 0.05$. 
As $l_\mathrm{av}$ and $n_\mathrm{av}$ scale linearly at the critical point, the Harris criterion under minimal assumptions\citep{Chayes1986, Chandran2015} now translates into a bound $\nu\geq\frac{2}{d+2}$, consistent with $\nu=1$.
Note that some recents studies\citep{Gray2018, Lenarcic2018} have found higher values of the exponent compatible with the standard Harris criterion.
The quality of the collapse for different values of $\Delta$ is unchanged, and the obtained exponents agree within the precision of our numerical analysis. 
This result therefore strongly suggests an extensive length of the minimal clusters at the phase transition.\\

At the phase transition, the minimal normalized mutual information $i_\text{min}$ averages to $0.11$ for $L=16$, significantly lower than the thermal value, and decreases with system size.
Our data is compatible with sublinear entanglement for this minimal bipartition: close to $W=3.8$,  $\langle i_\text{min} \rangle$ scales as $L^{-0.5\pm0.2}$, which is compatible with a power law scaling for the entropy of the minimizing cut $S(l_\Aa)\propto \sqrt{l_\Aa}$. 
Larger systems would be required to obtain a more precise bound on this exponent, since the minimal mutual information is subject to strong finite-size effects. 
Indeed, a finite-size-induced cross-over in the distribution of $i_{\min}$ is observed  at much lower values of the disorder than the critical point, while a decent collapse for intermediate values of $i_{\min}$ is obtained in the critical region (see Appendix \ref{app-miniClusters} for more details). 
Recent computation\citep{Pietracaprina2018} of the average entanglement in large-scale exact diagonalization nonetheless found it to be linear at the critical point.
Though one cannot exclude that sparse thermal clusters do exist in the thermodynamic limit due to the small size of the considered system, the weak sparsity of the clusters at all disorder strengths, as shown in Fig. \ref{fig:Continuity}\hyperref[fig:Continuity]{(c-d)}, strongly implies that this scenario does not occur. 
Indeed the difference between range and number of sites of clusters stays negligibly small, both in average and for the longest cluster of each chain. 
Though some clusters are indeed disconnected, the difference between range and cluster length is essentially never larger than one,  corresponding to a strong local impurity. 

\begin{figure}
\subfloat{\includegraphics[width=1\columnwidth]{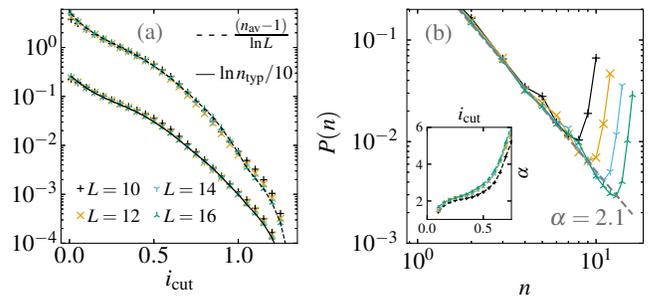}}
\caption{ Left panel shows the average length rescaled  with logarithm of system size $(n_\mathrm{av}-1)/\ln \ L$  (top dotted line) and the logarithm of the typical length $\ln(n_\mathrm{typ})/10$ (bottom full line) at the critical point as a function of the cut-off $i_\mathrm{cut}$. 
The logarithm of the typical length has been rescaled for convenience, and the lines are guides to the eye. 
We observe a slowly divergent average length of a cluster, while the typical length is itself constant. 
Right panel shows the distribution of the cluster length for $i_\mathrm{cut}=0.2$, with a decay as $l^{-\alpha}$, until finite-size effects kick in. 
The dotted line is the best power-law fit. 
All graphs are taken at the estimated critical point $W_c\approx 3.8$.
In inset, we show the evolution of $\alpha$ when varying the cut-off for $L=16$.
}
\label{fig:varied-cutoff-2}
\end{figure}

\begin{figure}[ht]
\begin{center}
\includegraphics[width=\linewidth]{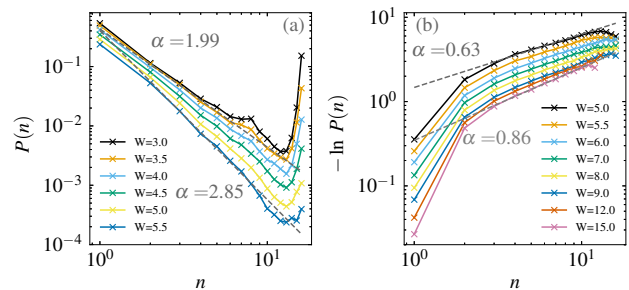}
\end{center}
\caption{Variation of the distribution of cluster lengths as a function of disorder strength, for $i_\mathrm{cut}=0.2$ (similar results are obtained for $0.1\leq i_\mathrm{cut} \leq 0.7$)  and for $L=16$. 
$(a)$ Around the estimated critical point, we observe a large region compatible with a power-law probability distribution $P(n)\propto l^{-\alpha}$, with exponents that vary slowly with disorder strength .
Note that a regression with $P(n)\propto\exp(- \beta \sqrt{l})$ also fits decently the intermediate cluster lengths. 
$(b)$ At larger disorder strengths, the distribution becomes a stretched exponential $P(n)\propto\exp(- \beta l^\alpha)$, with $\alpha$ going to $1$ as disorder increases.
}
\label{fig:probaDist-closeCritical}
\end{figure}

The picture of a weakly localized critical point is reinforced by an average and maximal cluster lengths that are both sublinear. 
We observe that the average mean length $n_\mathrm{av}$ diverges as $\ln L$ while the typical mean length $n_{\mathrm{typ}}=\exp \langle [ \ln n  ] \rangle$ is size-independent for $i_{\text{cut}}\geq 0.2$. 
This can be confirmed and understood by looking at the distribution of the cluster lengths, which follow a power-law until finite-size effects kick in, as shown in Fig.~\ref{fig:varied-cutoff-2}. 
The exponent increases with $i_\text{cut}$, but the power-law fitting is truly meaningful only for $0.15\leq i_\mathrm{cut}\leq 0.5$. 
Below this range, most systems are considered fully thermal, likely due to finite-size effects. 
Above it, clusters larger than one site become rare events, and we would require a larger number of realizations to verify that the distribution is still a power-law.
Interestingly, the exponent obtained in this meaningful regime  is close to the prediction obtained in Ref. \onlinecite{Goremykina2018}: $\alpha=2$.
We further observe a large window of disorder strengths wherein the  distribution is compatible with a power-law distribution, as shown in Fig. \ref{fig:probaDist-closeCritical}. 
The power-law distribution survives even when studying the general fitting form $P(n)\propto n^{-\alpha} \exp(-\beta n^\gamma)$: in a significant region around the critical point, we obtain $\gamma\approx 0$.
At a fixed $i_\mathrm{cut}$, exponents depend only weakly on the disorder strength.
While the finite-size effects at the tails of the distributions tend to diminish when increasing disorder strength, it remains unclear whether this indicates an actually higher value of the critical disorder, or if it is an effect of the cross-over to a stretched exponential behaviour.
The existence of such an intermediate region with power-law decay before a cross-over to stretched exponential was argued in Ref. \onlinecite{Dumitrescu2018}.
Indeed, at even stronger disorder, the power-law becomes a stretched exponential decay before turning into a simple exponential at very high disorder.
The exponent of the stretched exponential varies continuously with disorder strength and, close to the critical point, takes a value close to the one obtained by Ref. \onlinecite{Zhang2016} for correlation function decay in the localized fractal phase.

Finally, a qualitative comparison of the behaviour of the different characteristic lengths, both in a single state or averaged over disorder realizations, implies that critical states have a multiscale (or fractal) entanglement structure, as proposed in Refs.~\onlinecite{Serbyn2016-2, Zhang2016, Geraedts2016, Serbyn2017, Goremykina2018, Dumitrescu2018}. 
When increasing the entanglement cut-off for a given state, clusters progressively subdivide into smaller and smaller subclusters, instead of simply directly breaking down to single-site elements as in the ergodic phase. Some typical examples are presented in  Appendix \ref{app-hierarchy}.

\section{Discussion and conclusions}
We introduced a recursive numerical algorithm that allows the identification of entanglement clusters in wave functions. 
We used this algorithm to obtain and characterize the entanglement clusters in eigenstates of the XXZ Hamiltonian with a random field, which undergoes many-body localization at strong disorder.
The cluster structure in the ergodic and MBL phases follow expectations: large many-body entanglement in the ergodic phase, which turns into a set of weakly entangled spins, with a few resonant pairs in the localized phase. 
As a general rule, clusters are formed by neighbouring sites, with occurrences of ``holes" rare, even at the transition.
Comparison with random states and analysis of the sparsity and structure of the clusters suggest a weakly localized critical point, with  fractal or multiscale states. 
The distribution of cluster sizes at the critical point follows a power law with exponent close to $2$.
This value corresponds to the one obtained by mapping the MBL phase transition to a Kosterlitz-Thouless type of transition\citep{Goremykina2018, Dumitrescu2018}.
The power-law distribution of clusters seems to extend into the many-body localized phase, with a varying power eventually becoming a stretched exponential, and finally an exponential deep in the localized phase.
Though the spin-chain is typically localized at the critical point, some rare extended regions survive, which leads to a logarithmic growth of the average cluster size.
Our results are qualitatively consistent with the results of recent RG proposals\citep{Dumitrescu2017, Goremykina2018, Thiery2017, Thiery-2017-long},which are mostly phenomenological approaches.
The different scalings of the average and typical length of the thermal clusters was predicted by Ref.~\onlinecite{Goremykina2018}, while a non-universal power-law decay\citep{Dumitrescu2017} followed by a stretched exponential behaviour was proposed by Refs.~\onlinecite{Thiery2017, Thiery-2017-long, Dumitrescu2018}.
Conversely, Ref.~\onlinecite{Khemani2017} argued based on numerical arguments for the existence of a sparse thermal backbone.
More precisely, by analyzing single-site entropies they inferred the existence of contiguous sets of high-entropy sites, which formed local clusters, and then argued for (thermal) entanglement between these different clusters.
Our numerical analysis indicates that these clusters should instead be considered independent.
Nonetheless, both larger system sizes on the numerical side and a more microscopic approach on the renormalization side would be required for a careful validation.
These new results should put strong constraints on similar renormalization schemes.

An exciting perspective, though numerically challenging, would be to extend this study to larger size system using matrix-product states\citep{Khemani2016, Yu2016, Lim2016, Devakul2016, Serbyn2016, Doggen2018} (from the localized side of the phase transition) or machine-learning based approaches\citep{Carleo2017}. 
The study of these clusters close to the phase transition may give a clearer answer on the nature of the phase transition, and the existence of a potential intermediate phase of a non-ergodic metal.\citep{Altshuler1997, Serbyn2017, Torres2017, Znidaric2016, Luitz2016, Luitz:2017cp, Dumitrescu2018}
Extension of our results to higher dimensions, in particular to two-dimensions where tensor networks have been used to describe the MBL phase\citep{Wahl2017}, is also relevant.
A recent work on the mapping between excited localized eigenstates and ground states of disordered Hamiltonians\citep{Dupont2018} gives another promising means to numerically explore this problem. 
Properties of the entanglement clusters may shed light on the properties and stability of the many-body localized phase in this context.

\acknowledgments
This work was supported by the ERC Starting Grant No.~679722 and the Knut and Alice Wallenberg Foundation 2013-0093. 
S.B. acknowledges support from DST, India, through Ramanujan Fellowship Grant No.~SB/S2/RJN-128/2016.
This research was supported in part by the National Science Foundation under Grant No.~NSF PHY-1748958.

\appendix

\counterwithin{figure}{section}
\renewcommand\thefigure{\thesection\arabic{figure}}    
\section{Verifying algorithm validity}\label{app-Validity}
The algorithm we introduced in the main text is based on a couple of approximations that at first sight may seem uncontrolled. 
In this appendix we demonstrate that it correctly identifies the structure of randomly generated wave functions, and that for small systems where we can calculate all bipartition the approximation of continuous clusters does not affect the results.

\subsection{Randomly generated wave functions}
\begin{figure}[tb]
\includegraphics[width=\linewidth]{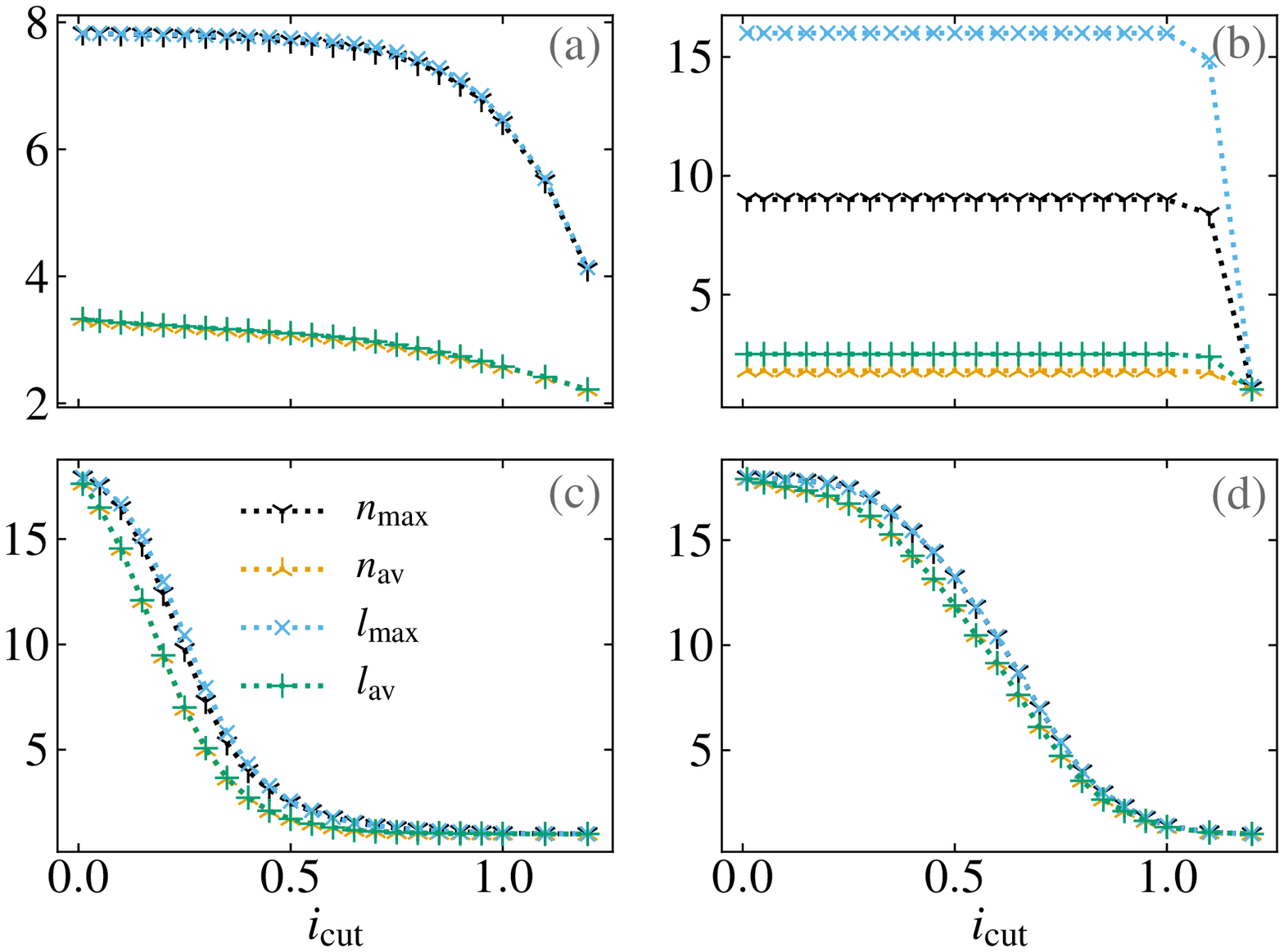}
\caption{The different characteristics lengths for random states of length $L=18$, generated according the structures described in Fig. \ref{fig:clusters-RG}. 
In (a), ergodic clusters survive high entanglement cut-off, while the localized one are broken down quickly. 
In (b), the sparse ergodic cluster leads to a significant discrepancy between maximum range and maximum size. 
In (c-d), multiscale entanglement leads to a progressive separation of clusters into smaller ones, and the difference between average and maximum cluster length is much smaller.
The difference between (c) and (d) lies in the probability distribution of wave function coefficients; see text for further details on how the random states are generated.
}
\label{fig:testSynthetic}
\end{figure}

We first show that our algorithm distinguishes between the entanglement structures described in Fig.~\ref{fig:clusters-RG}, by applying it to randomly generated wave functions; the results are shown in Fig.~\ref{fig:testSynthetic}.

The general process to create the random states is as follows:  we first randomly select the sets of spins that are in independent thermal or localized clusters. 
Localized clusters are broken down to single-site clusters. 
Due to the $U(1)$ symmetry, we assign a fixed magnetization to each cluster, depending on its size, so that the complete system stays in the total $\sigma^z=0$ sector. 
Then, for each cluster of size $n$ and with magnetization $m$, we generate a uniformly distributed vector in the corresponding Hilbert space, which is the hypersphere of $\mathbb{C}^{\binom{n}{m+n/2}}$. 
The global wave function is then simply the (properly reordered) tensor product of these wave functions. 
Noise can be added by taking superpositions of such vectors, that can either be different realizations of the same cluster decomposition, or have entirely different structures.
Though our results in Fig.~\ref{fig:testSynthetic} are presented without such noise, we verified they survive the added disorder.

In Fig.~\ref{fig:testSynthetic}\hyperref[fig:testSynthetic]{(a)}, we create continuous independent clusters, such that the probability for a cluster to be of length $n$ decays as $n^{-2}$, to mimic the distribution observed at the critical point.
The alternating ergodic and MBL structures are easily identified as the large ergodic clusters are still present at high values of $i_{\text{cut}}$, while the localized ones are reduced to a set of single-site clusters, resulting in a large difference between the average and maximum cluster sizes.
We also performed a similar analysis by alternating thermal clusters taken as blocks of varying length $n$, separated by sets of localized spins treated as single-site clusters, without significant differences.

For Fig.~\ref{fig:testSynthetic}\hyperref[fig:testSynthetic]{(b)}, we randomly select a set of $9$ spins that therefore spans most of the chain, and take the rest to be localized. 
The sparse ergodic cluster in this case leads to a significant difference between range and number of sites of the largest clusters. 
Varying the length of the sparse thermal cluster or introducing small thermal clusters does not affect the qualitative picture.

Finally, for Fig.~\ref{fig:testSynthetic}\hyperref[fig:testSynthetic]{(c-d)}, we create a multi-layered state in the following way. 
Let $\psi([s_1, s_2, ...], [m_1, m_2, ...])$ be the function that generates a random state associated to the cluster decomposition into sets $s_j$ with magnetization $m_j$. 
Then, the states we study here are given by:
\begin{equation}
\Ket{\psi(m_1, ..., m_L)} \propto \sum\limits_{j=0}^{\lceil  \ln_2 L \rceil} a_j \Ket{\psi_j} 
\end{equation}
with $a_j$ uniformly distributed random variables taken between either $0$ and $\frac{1}{j+1}$ [Fig.~\ref{fig:testSynthetic}\hyperref[fig:testSynthetic]{(c)}] or between $0$ and $1$ [Fig.~\ref{fig:testSynthetic}\hyperref[fig:testSynthetic]{(d)}], and
\begin{equation}	
\Ket{\psi_j}=\psi([ [1:2^j], [1+2^j:2^{j+1}], ...], [\sum\limits_{k=1}^{2^j} m_k, ...]), 
\end{equation}
where $m_k=\pm\frac{1}{2}$ is the local magnetization. 
In other words, we create wave functions that are superpositions of random states organized as follows: $\Ket{\psi_1}$ is a product state, while in $\Ket{\psi_{2 \leq j < \lceil  \ln_2 L \rceil }}$ spins are entangled in consecutive sets of $2^{j-1}$ sites, each set being in a product state with the others.  
Finally, $\Ket{\psi_{\lceil  \ln_2 L \rceil}}$ is a random thermal state. Magnetization is chosen such that the $U(1)$ symmetry is respected at all scales, and that the different states are coherent with one another. Finally, we attribute a random weight to each $\Ket{\psi_j}$.
A simple minimal example for $L=4$ is:
\begin{align}
\Ket{\psi(\frac{1}{2}, -\frac{1}{2}, -\frac{1}{2}, \frac{1}{2})} &\propto \Ket{\uparrow \downarrow \downarrow \uparrow} + \frac{1}{2}(\Ket{\uparrow \downarrow}-\Ket{\downarrow \uparrow})(\Ket{\uparrow \downarrow}+\Ket{\downarrow \uparrow})\nonumber \\ &+ \frac{1}{3} (\Ket{\uparrow \uparrow \downarrow \downarrow}+\Ket{\downarrow \downarrow\uparrow \uparrow}-\Ket{\downarrow \uparrow \downarrow\uparrow}+\Ket{\uparrow \downarrow  \uparrow \downarrow})
\end{align}
This hierarchy leads to a different dependence of $n$ and $l$ on $i_{\text{cut}}$, with an initial fast decay followed by a slowing down when all weakly entangled clusters have broken down.
In the localized phase, the decay is generally super-exponential with  $i_{\text{cut}}$, while  multiscale wave functions can have exponential or stretched exponential behaviour, depending on the exact choice of the entanglement structure. 
When studying a single wave function, the clusters are progressively fragmented at very different cut-off strengths.

\begin{figure}[tb!]
\begin{center}
\includegraphics[width=\linewidth]{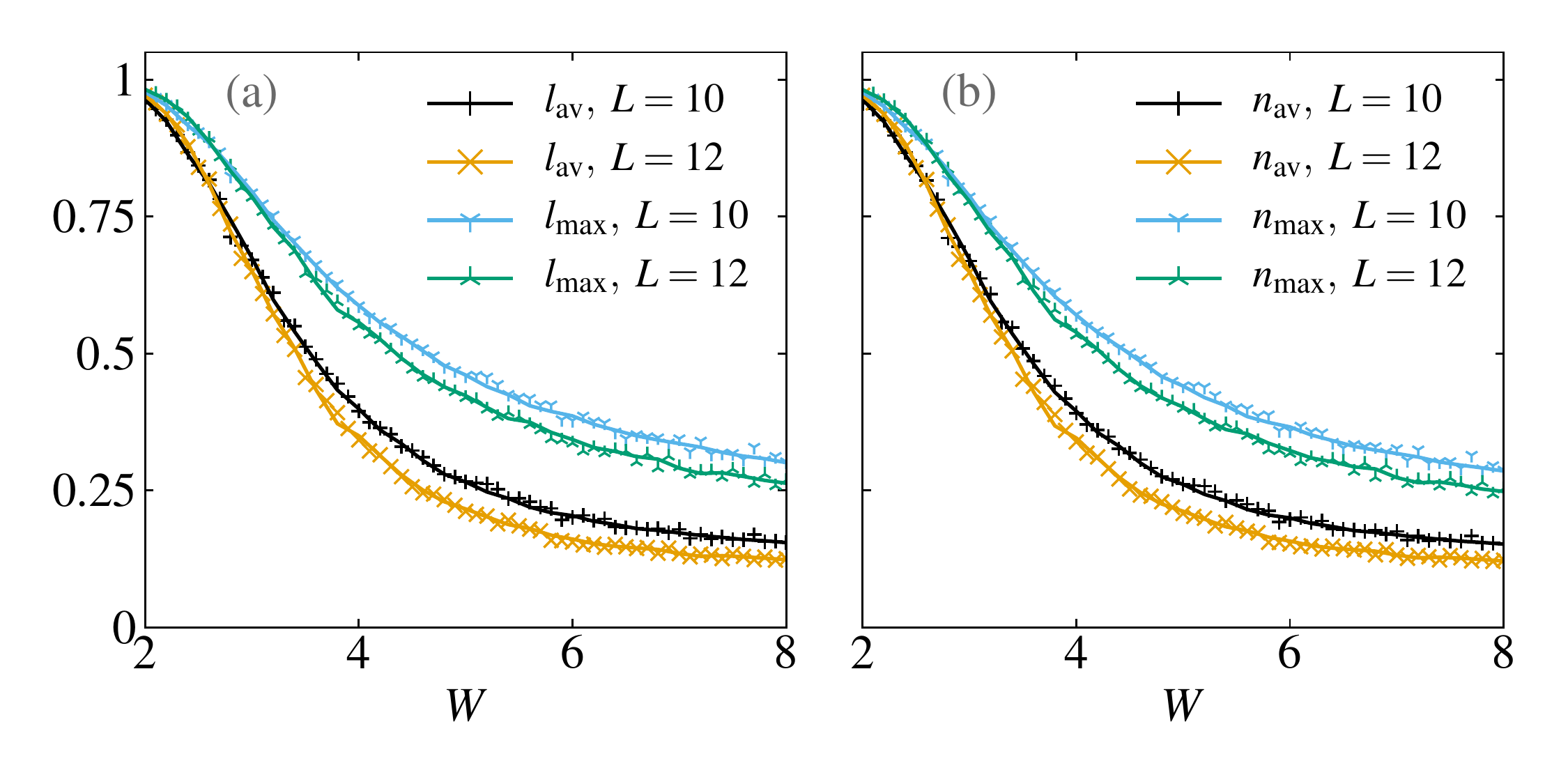}
\end{center}
\caption{Comparison of the different length scales---range in (a) and number of sites in (b)---between simulations where we evaluate the entropy on all bipartitions of the system (solid lines), and simulations where we only consider contiguous partitions in each subset (symbols). We take $i_{\text{cut}}=0.2$. We observe a good agreement for all disorder strengths and all entanglement cut-offs we consider.
}
\label{fig:ComparisonComplete}
\end{figure}

\subsection{Including noncontinuous bipartitions}
Our algorithm assumes that intertwined thermal clusters are inexistent, or at best, rare and irrelevant events, as we limit ourselves to continuous subsystems in our recursive bipartitioning of the system.
We checked on small systems, where all bipartitions can be computed, the validity of this approximation.
Figure \ref{fig:ComparisonComplete} presents a comparison between the different length scales we obtain with our algorithm and the case when we compute the mutual information of all bipartitions.
No significant differences is observed.

\section{Cluster decomposition example}\label{app-Decompo}

\begin{figure}[tb!]
\begin{center}
\begin{minipage}{0.4\linewidth}
\subfloat{\includegraphics[width=\linewidth]{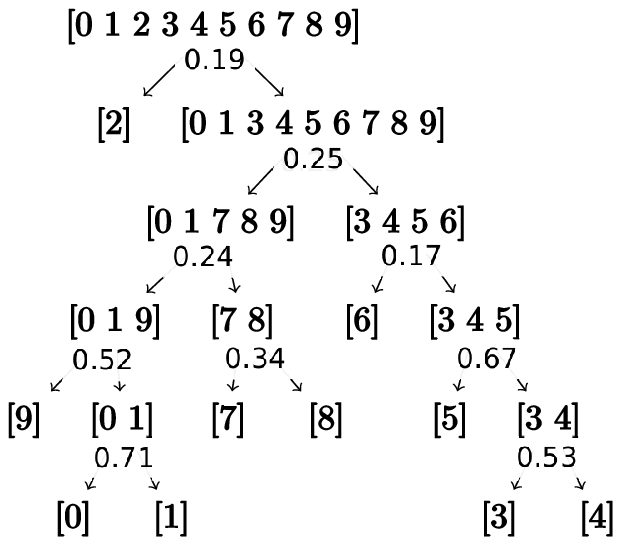}}\hspace{0.3cm}
\end{minipage}\hspace{0.2cm}
\begin{minipage}{0.5\linewidth}
\subfloat{\includegraphics[width=\linewidth]{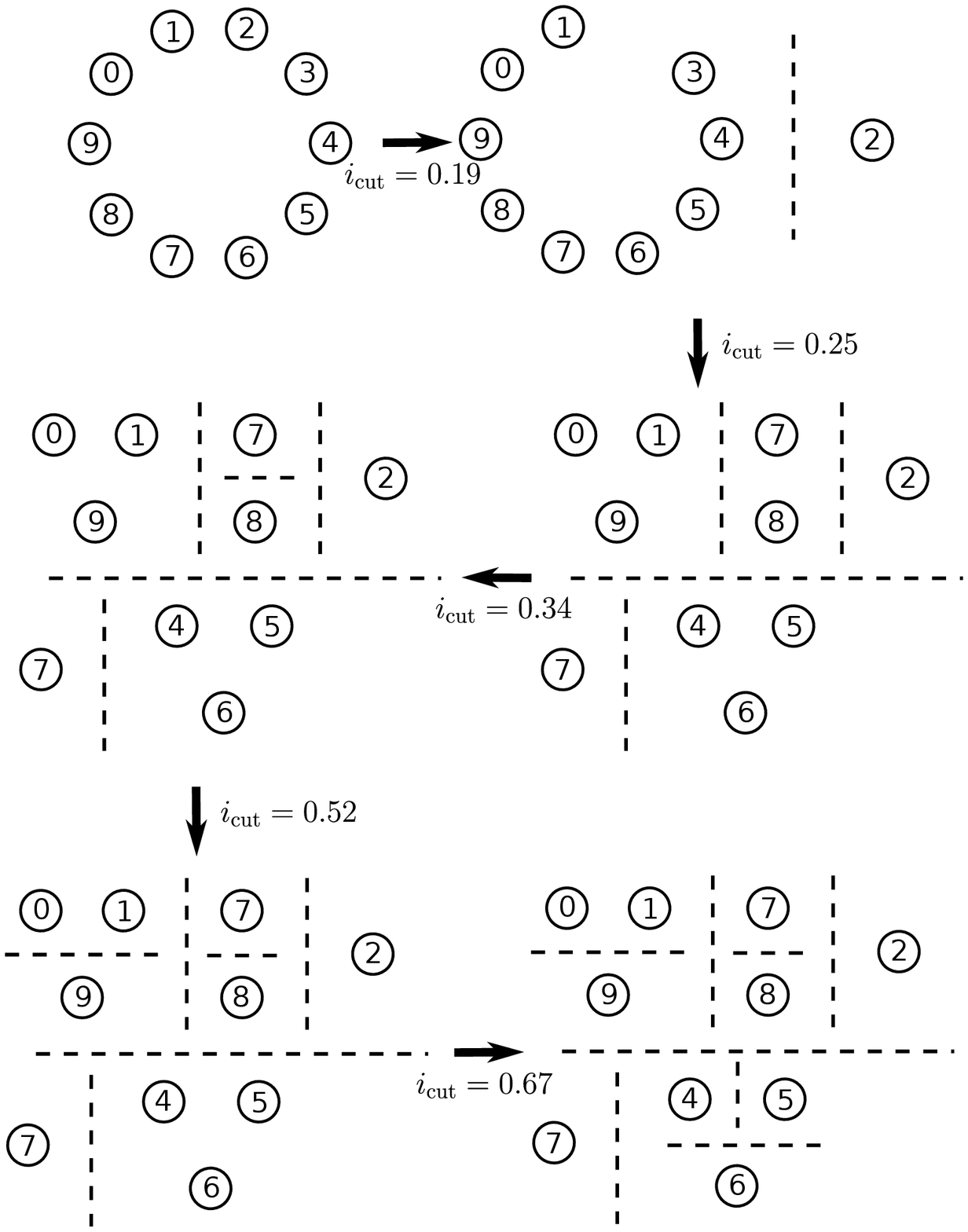}}\\
\end{minipage}
\end{center}
\caption{Example of a decomposition into clusters by our algorithm (with $L=10$ and $W=3.7$). 
On the left, the binary tree we obtain, with each node corresponding to a set of spins labelled by the sites it includes (here from 0 to 9).
Sets are divided into subsets following the edges of the tree. 
The label on the edges corresponds to the normalized mutual information between its two descendants. 
On the right, a schematic picture of the progressive appearance of clusters when varying the mutual information cut-off $i_{\text{cut}}$, applied to the binary tree on the left. 
The indexed circles correspond to sites, and the cluster decomposition is represented by the dashed lines. 
Each arrow corresponds to a change in the structure at some precise value of the mutual information cut-off. 
As a concrete example, we obtain the second decomposition for $0.19 \leq i_{\text{cut}}<0.25$.}
\label{fig:exampleDecomposition}
\end{figure}

We provide in Fig.~\ref{fig:exampleDecomposition} an example of a decomposition obtained by our algorithm, when we vary the mutual information cut-off introduced in the main text. 
The state we study was obtained for $L=10$ and $W=3.7$. 

\section{Complementary numerical results}
In this appendix, we provide additional numerical results that support and supplement claims made in the main text.

\begin{figure}[tb]
\begin{center}
\includegraphics[width=\linewidth]{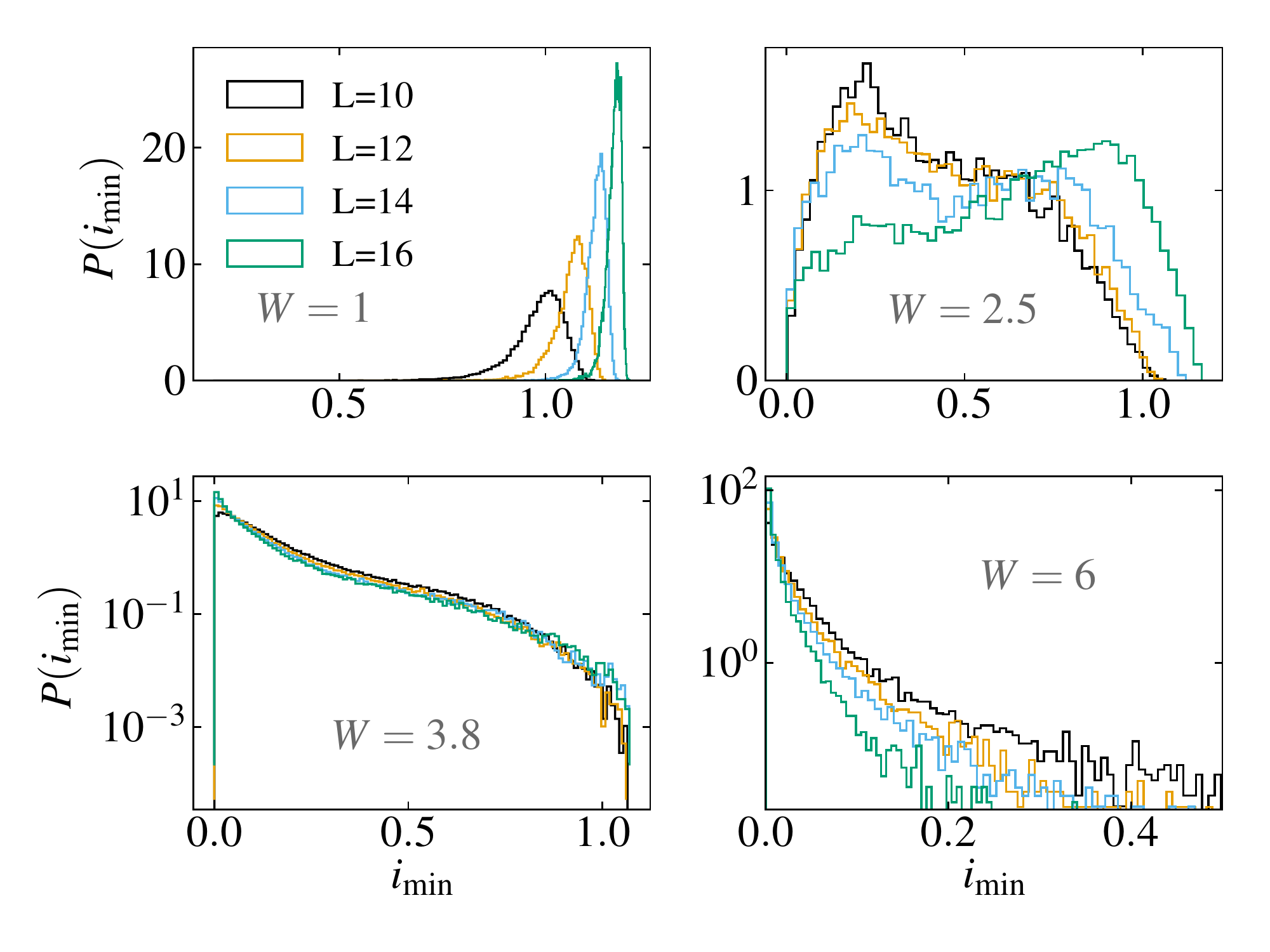}
\end{center}
\caption{Probability distribution of the minimal mutual information for different disorder strengths and system sizes. 
Deep in the ergodic phase, the probability distribution peaks around the (renormalized) thermal entropy. 
For $W=2.5$, strong finite-size effects are visible, which correspond to the onset of the transition in small systems. 
At the phase transition, we observe a decent collapse at intermediate values, though the low-entanglement realizations vastly differ. 
The average minimal mutual information $\langle i_{\min} \rangle$ decreases with system size as $L^{-0.5 \pm 0.2}$.
}
\label{fig:miniEntropy}
\end{figure}

\subsection{Minimal cluster properties}\label{app-miniClusters}

Figure \ref{fig:miniEntropy} illustrates the strong finite-size scaling apparent in the minimal mutual information distribution.
For $W=1$, the distribution admits a clear thermal peak, which increases with system size and converges to the thermodynamic thermal value. 
For $W=6$, the system has become localized, and the distribution now peaks at $i_{\min}=0$, with a suppressed tail when system size increases.
In between, we observe a cross-over, with an intermediate bimodal distribution at low value of the disorder strength $W=2.5$.
Similar bimodal distributions were observed in other studies on entanglement distributions\citep{Yu:2016gb, Dumitrescu2017, Khemani2017}, but here this bimodality is seemingly only a transient finite-size effect.
Close to the estimated phase transition, the distributions collapse decently at intermediate values of $i_{\min}$.
Scaling analysis on the characteristic lengths of the minimal clusters also shows a good scaling behavior, as has been shown in the main text.

\begin{figure}[ht]
\begin{center}
\includegraphics[width=\linewidth]{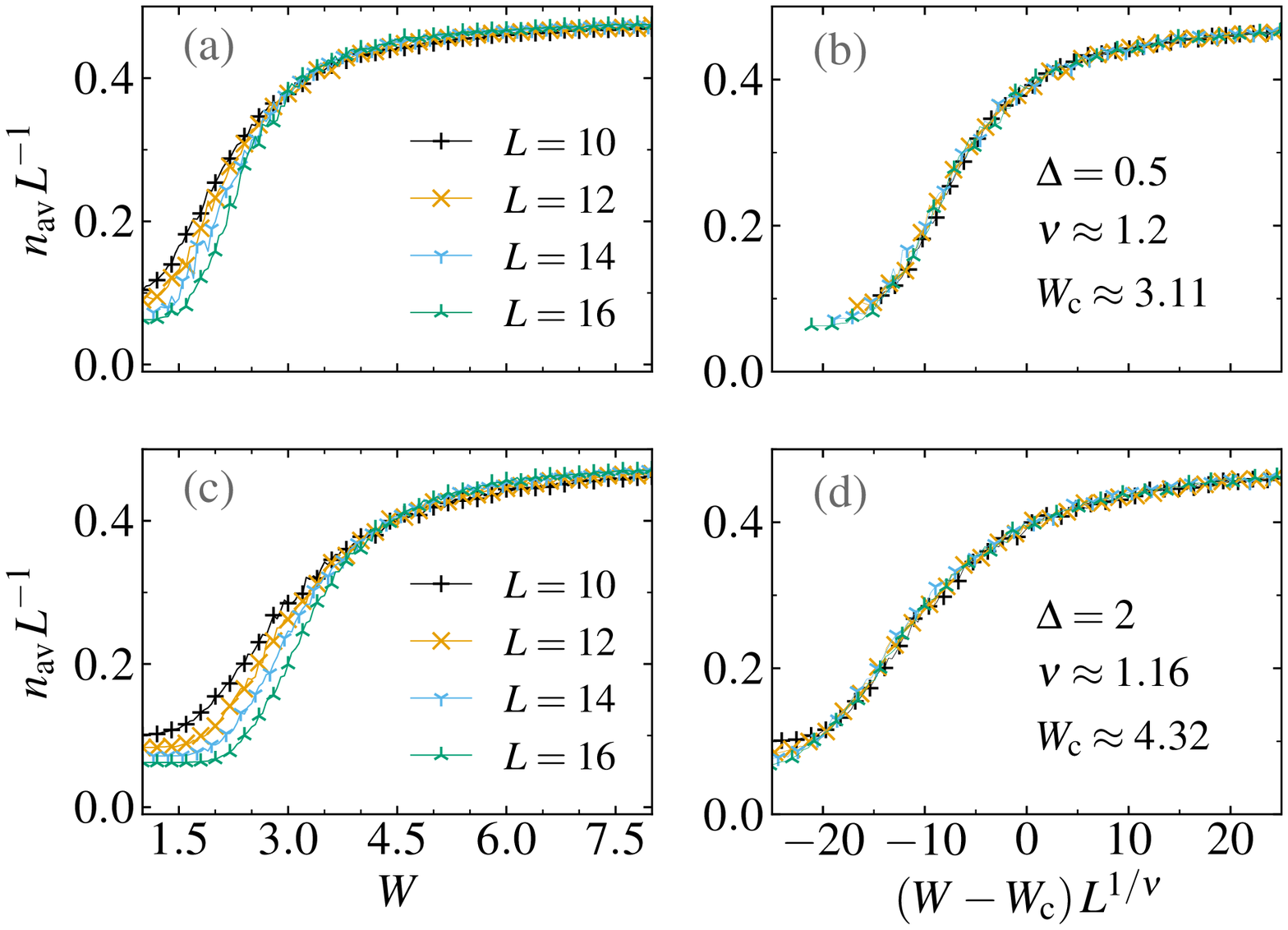}
\end{center}
\caption{Statistical average of the mean length of the minimal clusters for two values of the interaction strength $\Delta$, away from the usual Heisenberg limit. 
(a-b) are taken for $\Delta=0.5$, while (c-d) correspond to $\Delta=2$. 
(b) and (d) are fits to the universal scaling form $n_\mathrm{av}=L^{-1} f((W-W_c)L^{1/\nu})$. 
The exponents obtained are similar to the one for $\Delta=1$.}
\label{fig:miniClusters}
\end{figure}

Figure \ref{fig:miniClusters} presents the scaling of the average length of the minimal clusters for different interaction strengths $\Delta=0.5$ and $\Delta=2$. 
In both cases, this quantity acts as an order parameter, with a collapse similar to the one observed for $\Delta=1$.  
The critical exponents qualitatively agree with the one observed in the Heisenberg limit.

\begin{figure*}[tb!]
\begin{center}
\includegraphics[width=\linewidth]{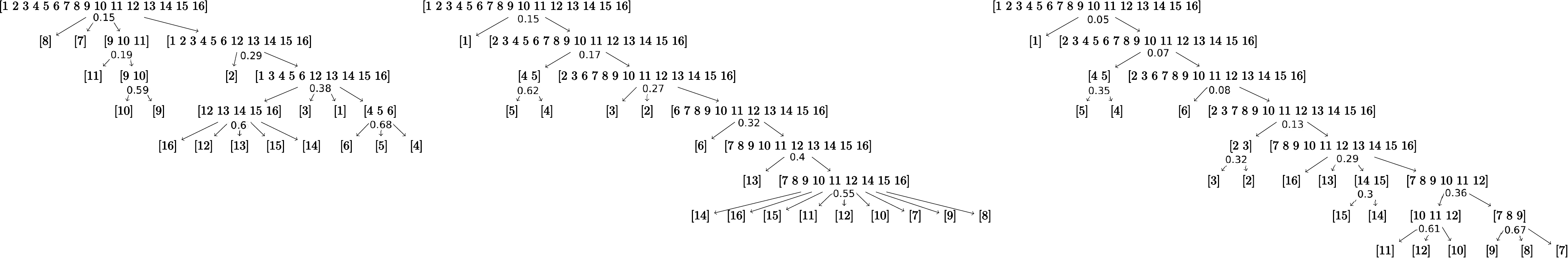}
\end{center}
\caption{Three typical examples of the hierarchical multiscale structure obtained for states close to the critical point, with $W=3.7$ and $L=16$.  
Each node corresponds to a cluster, labelled by the sites it includes (here from 1 to 16). 
The label on the edges corresponds to the mutual information cut-off $i_{\text{cut}}$ required to break it into its descendants. 
For ease of visualization, the data is no longer a binary tree; we have only kept the nodes that appear at the given value of the cut-off, and removed all that are less entangled than their parents (and therefore never appear as a proper cluster). 
As is readily seen, thermal subclusters (strongly entangled clusters that break into single-site clusters) are present. 
Clusters can themselves be entangled at larger scales, with a weaker internal entanglement ($[1\ 3\ 4\ 5\ 6\ 12\ 13\ 14\ 15\ 16]$  in the first example, or $[7\ 8\ 9\ 10\ 11\ 12]$ in  the third one are good examples of such structures).
}
\label{fig:structure}
\end{figure*}

\subsection{Hierarchical structure at the critical point}\label{app-hierarchy}
Finally, we provide in Fig.~\ref{fig:structure} explicit examples of the typical entanglement structures observed in states close to or at the critical disorder strength. 
Strongly entangled small subsets of the system become entangled with other subsets at larger scales. 
This multiscale structure is strongly indicative of Griffith or fractal states.

\bibliography{MBL}

\end{document}